\documentclass[sigconf]{acmart}

\AtBeginDocument{%
  }

\setcopyright{acmlicensed}

\copyrightyear{2025}
\acmYear{2025}
\acmConference[UIST '25]{The 38th Annual ACM Symposium on User Interface Software and Technology}{September 28-October 1, 2025}{Busan, Republic of Korea}
\acmBooktitle{The 38th Annual ACM Symposium on User Interface Software and Technology (UIST '25), September 28-October 1, 2025, Busan, Republic of Korea}
\acmDOI{10.1145/3746059.3747730}
\acmISBN{979-8-4007-2037-6/2025/09}

\usepackage{xspace}
\newcommand{\oursystem}{Samosa\xspace}
\newcommand{\SamosaGeoOnly}{Samosa-GeoOnly\xspace}
\newcommand{\SamosaMatOnly}{Samosa-MatOnly\xspace}
\newcommand{\SamosaAEOnly}{Samosa-AEOnly\xspace}
\newcommand{\SamosaOurs}{Samosa (Ours)\xspace}
\newcommand{\NonAdaptive}{Non-Adaptive\xspace}

\newcommand{\Naturalness}{\textit{Naturalness}\xspace}
\newcommand{\Clarity}{\textit{Clarity}\xspace}
\newcommand{\Externalization}{\textit{Externalization}\xspace}

\newcommand{\rt}{\textit{RT60}\xspace}
\newcommand{\edt}{\textit{EDT}\xspace}

\newcommand{\ourheadset}{customized XR headset equipped with the Snapdragon XR2+ Gen 2 chipset \cite{qualcomm:xr2plusgen2_brief}\xspace}
\newcommand{\ourcompany}{\textit{Google}\xspace}
\newcommand{\avpfacetime}{\textit{AVP FaceTime}\xspace}
\newcommand{\imagereverb}{\textit{Image2Reverb}\xspace}
\newcommand{\avrir}{\textit{AV-RIR}\xspace}

\newcommand{\eg}{\emph{e.g.,}\xspace}
\newcommand{\dquote}[1]{``#1''\xspace}
\newcommand{\squote}[1]{`#1'\xspace}

\usepackage{amsmath}
\usepackage{enumitem}
\usepackage{subfigure}
\usepackage{listings}
\usepackage{xcolor}
\usepackage{caption}
\usepackage{hyperref}
\usepackage{threeparttable}
\usepackage{booktabs}
\usepackage{url}
\usepackage{fancyvrb}

\definecolor{codegreen}{rgb}{0,0.6,0}
\definecolor{codegray}{rgb}{0.5,0.5,0.5}
\definecolor{codepurple}{rgb}{0.58,0,0.82}
\definecolor{backcolour}{rgb}{0.95,0.95,0.92} 
\lstdefinestyle{mystyle}{
    backgroundcolor=\color{backcolour},
    commentstyle=\color{codegreen},
    keywordstyle=\color{magenta}, 
    numberstyle=\tiny\color{codegray},
    stringstyle=\color{codepurple},
    basicstyle=\ttfamily\small, 
    breakatwhitespace=false,
    breaklines=true,                 
    captionpos=b,                    
    keepspaces=true,
    numbers=left,                    
    numbersep=5pt,
    showspaces=false,
    showstringspaces=false,
    showtabs=false,
    tabsize=2,
    frame=single,                    
    rulecolor=\color{black},         
    title=\lstname                   
}
\newif \ifdraft \drafttrue   

\newif \ifhighlight \highlighttrue    
\newif \ifhighlight \highlightfalse    

\newcommand{\change}[1]{#1}

\begin{document}

\title{Enhancing XR Auditory Realism via Scene-Aware Multimodal Acoustic Rendering}


\settopmatter{authorsperrow=4}

\author{Tianyu Xu}
\orcid{0009-0009-9135-6080}
\affiliation{%
  \institution{Google}
  \city{Mountain View}
  \state{CA}
  \country{USA}
}
\email{tyx@google.com}

\author{Jihan Li}
\affiliation{%
  \institution{Google}
  \city{Mountain View}
  \state{CA}
  \country{USA}
}
\email{jihanli@google.com}

\author{Penghe Zu}
\affiliation{%
  \institution{Google}
  \city{Mountain View}
  \state{CA}
  \country{USA}
}
\email{penghezu@google.com}

\author{Pranav Sahay}
\affiliation{%
  \institution{Google}
  \city{Mountain View}
  \state{CA}
  \country{USA}
}
\email{prasahay@google.com}

\author{Maruchi Kim}
\affiliation{%
  \institution{Google}
  \city{Seattle}
  \state{WA}
  \country{USA}
}
\email{maruchi@google.com}

\author{Kwadwo Obeng\large-\LARGE Marnu}
\affiliation{%
  \institution{Google}
  \city{San Francisco}
  \state{CA}
  \country{USA}
}
\email{jackom@google.com}

\author{Farley Miller}
\affiliation{%
  \institution{Google}
  \city{Mountain View}
  \state{CA}
  \country{USA}
}
\email{farleymiller@gmail.com}

\author{Xun Qian}
\orcid{0000-0003-1976-7992}
\affiliation{%
  \institution{Google}
  \city{Mountain View}
  \state{CA}
  \country{USA}
}
\email{xunqian@google.com}

\author{Katrina Passarella}
\affiliation{%
  \institution{Google}
  \city{San Francisco}
  \state{CA}
  \country{USA}
}
\email{kpassarella@google.com}

\author{Mahitha Rachumalla}
\affiliation{%
  \institution{Google}
  \city{Mountain View}
  \state{CA}
  \country{USA}
}
\email{rachumalla@google.com}

\author{Rajeev Nongpiur}
\affiliation{%
  \institution{Google}
  \city{Mountain View}
  \state{CA}
  \country{USA}
}
\email{rnongpiur@google.com}

\author{D Shin}
\affiliation{%
  \institution{Google}
  \city{Mountain View}
  \state{CA}
  \country{USA}
}
\email{deshin@google.com}

\renewcommand{\shortauthors}{Xu et al.}
\begin{abstract}

In Extended Reality (XR), rendering sound that accurately simulates real-world acoustics is pivotal in creating lifelike and believable virtual experiences. However, existing XR spatial audio rendering methods often struggle with real-time adaptation to diverse physical scenes, causing mismatches between visual and auditory cues. This sensory conflict can induce cognitive dissonance, disrupting user immersion. Drawing from domain knowledge and prior works, we explore three key scene-based audio realism enhancement strategies in the design space: 1) room geometry approximation, 2) material segmentation, and 3) semantic-driven acoustic parameter estimation. We introduce \oursystem, a novel on-device system designed to render spatially accurate sound for XR by dynamically adapting to its physical environment. Leveraging multimodal Room Impulse Response (RIR) synthesis, \oursystem estimates scene acoustic properties—informed by room geometry, surface materials, and acoustic context—and subsequently renders highly realistic acoustics suitable for diverse settings. We validate our system through technical evaluation using acoustic metrics for RIR synthesis across various room configurations and sound types, alongside a human expert evaluation (N=12). Evaluation results demonstrate \oursystem feasibility and efficacy in enhancing XR audio realism.

\end{abstract}



\begin{teaserfigure}
  \includegraphics[width=\textwidth]{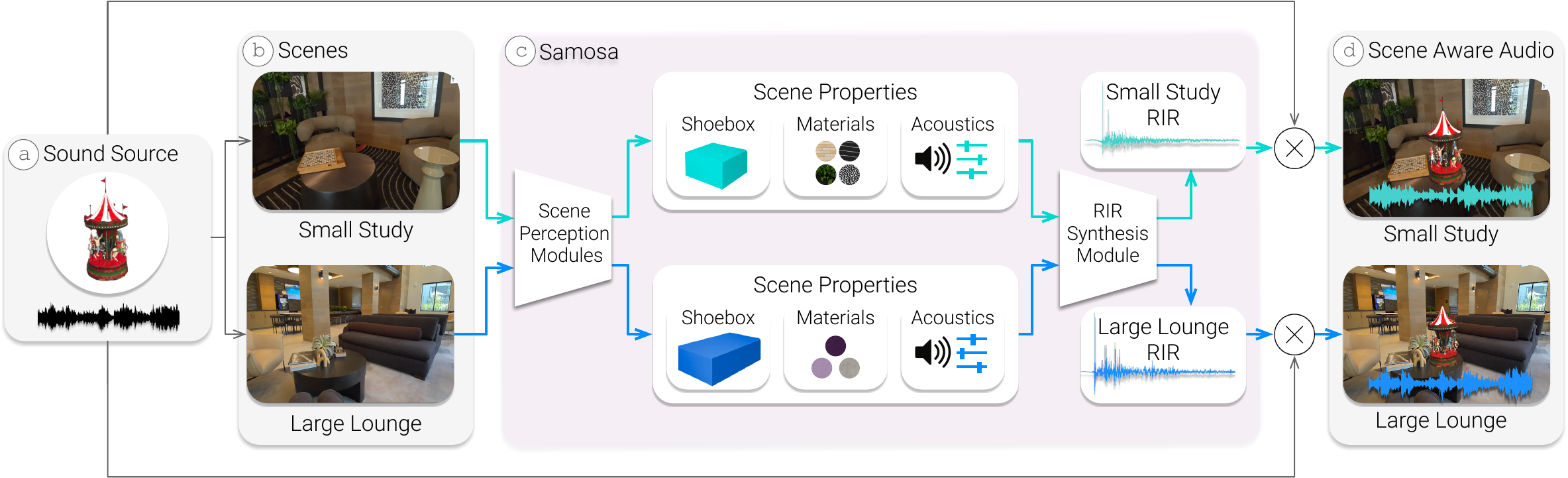}
  \caption{Overview of \oursystem adapting spatial audio to the user’s environment.
(a) A user in XR listens to a virtual sound source (music box).
(b) The system operates in diverse scenes, illustrated by a small study room and a large public lounge.
(c) \oursystem perceives scene geometry, materials, and acoustics in real time to synthesize distinct Room Impulse Responses (RIRs).
(d) Convolving ($\otimes$) the RIRs with input audio produces scene-aware sound that reflects the environment’s acoustic characteristics.}
  \label{fig:teaser}
\end{teaserfigure}
\newpage

\maketitle

\section{Introduction}
\label{sec:introduction}



Extended Reality (XR) technologies strive to create immersive experiences by seamlessly merging virtual content with the physical world \cite{milgram,steuer,azuma,jerald_2015}. Within these experiences, audio plays a pivotal role: it provides crucial spatial cues and contributes significantly to the realism and believability of the virtual environment \cite{hendrix_1996,serafin_2016}. When auditory cues align coherently with visual information, the user's sense of presence is substantially strengthened \cite{hendrix_1995, hendrix_1996}.

\change{The central challenge is achieving this acoustic realism while satisfying the unique demands of modern XR environments. The audio must be perceptually convincing; for instance, an XR user expects to hear the sound of a virtual character's voice change realistically as they both walk from a small, carpeted office into a large, reverberant atrium. Similarly, a virtual object placed near a real-world glass window should sound bright and reflective, while the same object near a cloth curtain should sound muffled and absorbed. This level of realism necessitates real-time rendering to maintain synchronization with the user's movements, and a dynamic system, adapting \dquote{on-the-fly} to changing surroundings without explicit commands or pre-scans. Solving these issues is therefore crucial for achieving true implicit auditory interaction, where the environment's acoustics are so seamlessly integrated they become a natural part of the user's experience \cite{sonic_interaction_design, ju_2015, cho_2024}.}

Prior work has struggled to meet all these requirements simultaneously. Geometric Acoustics (GA) methods present a difficult trade-off: high-fidelity simulations are too computationally expensive for mobile devices, while simplified, performant versions often fail to produce realistic audio \cite{tsingos_2002, kuttruff_2016}. On the other hand, learning-based approaches are powerful but face both prohibitive computational demands and a critical data bottleneck, requiring extensive, device-specific, and accurately paired visual-acoustic datasets that are impractical to collect at scale \cite{image2reverb_2021, avrir_2024, tong_2025}. Consequently, a critical research gap exists for an integrated, end-to-end system that can bridge the gap between acoustic fidelity and on-device efficiency.

Solving this challenge is now feasible due to the recent convergence of several key technological advancements. First, the rapid increase in on-device computational power and the inclusion of rich sensor suites (\eg depth, RGB cameras) in modern XR headsets provide a continuous stream of environmental data \cite{qualcomm:xr2plusgen2_brief, avp}. Second, a confluence of advances in machine learning has made complex environmental perception practical on-device. Efficient neural network architectures enable real-time scene understanding \cite{MobileNetV2, liu_2018, liu_2019, chen_2017_deeplabv3}, foundation models like SAM and LLMs allow for large-scale automated data labeling \cite{SAM, gemini}, and progress in cross-modal learning has demonstrated the potential to infer acoustic properties from visual data \cite{kim_2019, image2reverb_2021}. Finally, mature real-time audio engines provide a robust foundation for the final synthesis stage \cite{resonance}. 
The confluence of these technologies enables a new class of hybrid systems capable of using real-time environmental perception to drive dynamic, physically-plausible audio rendering, and therefore applying these technologies toward a cohesive, real-time system for XR devices still presents significant challenges.

To this end, we propose \oursystem (Scene-Aware Multimodal On-device Spatial Audio), a novel hybrid system for XR platforms that dynamically adapts audio rendering to the user's physical surroundings. Our approach intelligently combines a lightweight geometric acoustics engine with a set of efficient, learning-based perception models. \oursystem leverages existing XR sensors to estimate key scene characteristics—approximated geometry, surface materials, and overall acoustic context. By synthesizing plausible RIRs reflecting these characteristics on-the-fly, \oursystem renders audio that is acoustically coherent with the visual scene, enhancing XR auditory realism and immersion. 

Our primary contributions are:
\begin{enumerate}
    \item A novel multimodal scene-aware rendering approach for XR that balances a lightweight geometric acoustics engine with learning-based multimodal scene perception (geometry, material, and acoustic properties). This method enables dynamic, on-the-fly parameterization of the audio renderer to match the user's current physical environment.
    \item The design and implementation of \oursystem, the first end-to-end system that realizes this hybrid approach on XR devices. \oursystem integrates real-time multimodal sensing to synthesize plausible RIRs on resource-constrained XR hardware.
    \item A comprehensive evaluation including: (a) technical analysis of {\oursystem}'s acoustic accuracy and computational performance, and (b) a human expert study (N=12) demonstrating significantly enhanced perceived audio realism compared to baseline methods.
\end{enumerate}

\section{Related Work}
\label{sec:related_work}

Our work is positioned at the intersection of real-time spatial audio, scene-aware acoustic modeling, and on-device scene understanding. This section reviews prior research in these areas to identify the key challenges and opportunities that motivate our approach.

\subsection{Scene-Aware Audio Rendering for XR}
\change{Creating immersive audio in XR requires simulating how sound interacts with the user's environment. Early methods focused on listener-centric cues using Head-Related Transfer Functions (HRTFs) for directionality and generic reverberation for a basic sense of space \cite{wenzel_1993, kuttruff_2016}. However, these non-scene-aware techniques create a perceptual disconnect when the audio does not match the visual characteristics of the space (\eg a large hall sounding like a small room), which undermines realism and the user's sense of presence \cite{hendrix_1996}.}

\change{This challenge of creating meaningful and contextually appropriate auditory feedback is a central theme in Sonic Interaction Design (SID), which advocates for a more considered role for sound in the interaction loop \cite{sonic_interaction_design}. When a system can automatically generate this context-aware audio, it creates a form of implicit interaction, where the technology seamlessly adapts to the user's context without needing explicit commands \cite{ju_2015}. This has spurred active research into designing more effective and less ambiguous audio cues for XR interfaces \cite{cho_2024}. Thus, the broader goal of improving perceptual realism by dynamically matching the user's environment continues to be a key challenge in Human-Computer Interaction (HCI).}

To address this, modern research focuses on scene-aware rendering, which can be broadly categorized into three approaches.

\subsubsection{Physics-based Acoustics}
These methods simulate sound wave propagation using numerical techniques like FDTD or BEM \cite{yee_fdtd, bem}. While highly accurate in modeling complex phenomena like diffraction, their immense computational cost makes them unsuitable for real-time rendering in interactive XR applications.
\subsubsection{Geometric Acoustics}
Geometric Acoustics (GA) offers a computationally tractable approach by modeling sound as rays to efficiently calculate reflections and reverberation based on scene geometry \cite{kuttruff_2016, tsingos_2002}. Techniques like the Image Source Method (ISM) excel at calculating early reflections \cite{allen_1979}, and integrating frequency-dependent material properties is crucial for realism \cite{colombo_2022}. However, GA presents a difficult trade-off on mobile hardware: high-fidelity simulations are too computationally expensive for real-time use, while simplified, performant versions often fail to produce realistic audio. 

Examples of such efficient approximations include \dquote{shoebox} models, which assume simple rectangular rooms for fast calculation but are limited in complex spaces \cite{kuttruff_2016}. Another common approach is delay network-based reverberators, which use interconnected delay lines to simulate reverberation but can produce unnatural artifacts that reduce realism \cite{schroeder_1962, jot_1991}. Recent advancements in consumer XR hardware, such as the Apple Vision Pro, demonstrate the increasing feasibility of employing real-time audio ray tracing \cite{avp}. Such systems represent an important practical baseline for evaluating the performance and quality of real-time scene-aware GA rendering.


\subsubsection{Learning-based Acoustics}
Recent approaches use machine learning to predict acoustic properties or synthesize RIRs from sensor data. These models range from estimating semantic properties to inform acoustic models \cite{kim_2019}, to synthesizing RIRs or sound fields directly from images \cite{image2reverb_2021} or combined audio-visual streams \cite{avrir_2024}. 

Despite their power to capture nuanced acoustic effects, the primary limitation of these methods for interactive XR is computational cost. State-of-the-art models like Neural Acoustic Fields, for instance, can synthesize highly realistic room acoustics but demand substantial processing power that precludes real-time, on-device use \cite{luo_2023, lan_2024, tong_2025}. This computational burden makes it difficult to meet the strict low-latency and low-power requirements of mobile XR platforms.

Moreover, these models face a critical data bottleneck. Their accuracy and generalization to novel environments are highly dependent on vast, device-specific, and perfectly paired visual-acoustic datasets, which are impractical to collect and curate at scale. Despite these limitations for real-time systems, the output quality of these offline learning-based methods makes them valuable baselines.

\subsection{Real-Time Scene Understanding in XR}
Effective scene-aware audio is fundamentally dependent on the system's ability to perceive the user's environment in real time. This capability is driven by parallel advancements in on-device hardware and perception algorithms.

\subsubsection{Hardware and Sensor Platforms} Modern XR headsets are now equipped with powerful mobile chipsets and a rich suite of sensors, including stereoscopic cameras (RGB and/or monochrome) and depth sensors (using techniques like Time-of-Flight or Structured Light). This hardware, combined with data from Inertial Measurement Units (IMUs), provides a continuous stream of multimodal data. Furthermore, APIs like OpenXR offer standardized access to processed environmental data, such as detected planar surfaces and scene meshes, which form the raw input for perception algorithms \cite{openxr}.

\subsubsection{Algorithmic and Learning-based Perception} This sensor data is interpreted by a range of algorithms to build a comprehensive environmental model.
\begin{itemize}
\item Geometric Understanding: To understand the spatial layout, systems often employ sensor fusion techniques like SLAM (Simultaneous Localization and Mapping) to process IMU and camera data, providing a robust understanding of the user's position within the scene \cite{mur-artal_2015}. This is complemented by algorithms for planar reconstruction from depth data, which identify key surfaces like walls and floors essential for geometric acoustics \cite{liu_2018,liu_2019}.

\item Semantic and Material Understanding: Progress in cross-modal learning has shown that acoustically-relevant properties like surface materials can be inferred directly from visual data, avoiding the need for intrusive acoustic measurements \cite{kim_2019, image2reverb_2021, zhang_2016}. These tasks are powered by efficient deep learning architectures (e.g., DeepLabv3+, MobileNetV2) that are optimized for real-time, on-device inference \cite{chen_2017_deeplabv3, MobileNetV2}.

\item Foundation Models: The recent advent of foundation models like the Segment Anything Model (SAM) and large language models (LLMs) provides powerful new tools for zero-shot segmentation and large-scale automated data annotation, significantly mitigating the traditional data acquisition bottleneck that has hindered prior work \cite{SAM, gemini}.
\end{itemize}

Our work integrates these hardware and algorithmic advancements to build the comprehensive environmental model required for high-fidelity, dynamic acoustic rendering.

\section{Design Space}
\label{sec:design_space}

To create realistic spatial audio for XR applications, a system must simulate how sound interacts with the user's environment. A central challenge lies in identifying the most critical environmental information to capture and model, balancing potential gains in perceptual realism against the significant constraints of real-time performance on mobile devices. To justify our novel hybrid approach, we now explore this design space along the three primary dimensions of scene representation: \textbf{Geometry}, \textbf{Material}, and \textbf{Acoustic Properties}.

\subsection{Geometric Representation}
The physical shape and size of an environment fundamentally dictate sound propagation. The choice of geometric representation involves a trade-off between acoustic fidelity and computational cost. Options range from simplified shoebox models, which are fast but inaccurate in complex spaces \cite{kuttruff_2016}, to full 3D mesh reconstructions, which enable accurate simulation but are costly to acquire and process in real-time \cite{zipnerf_2023}. Planar surfaces (walls, floors) offer a powerful middle ground, capturing the most perceptually significant surfaces for early reflections with manageable complexity \cite{liu_2018, liu_2019}.

\subsection{Material Representation}
Surface materials determine how sound energy is absorbed and reflected, critically affecting reverberation timbre and decay. Ignoring materials compromises realism, while estimating detailed, frequency-dependent coefficients is often impractical for real-time estimation on mobile hardware. A balanced approach is to perform broad classification, categorizing materials into general types (\eg acoustically hard, soft) based on visual data to apply representative acoustic parameters, which provides a substantial boost in realism for a moderate cost \cite{davis_2020, colombo_2022}.

\subsection{Acoustic Representation}
Ultimately, the perceived environmental data must be translated into an acoustic context, which should represent the essential acoustic character of the scene, including parameters like Reverberation Time (RT60/EDT), Reflection Gain (Early Reflection Pattern), Reverberation Gain (or Direct-to-Reverberant Ratio), and Reverberation Brightness (Timbre) \cite{kuttruff_2016}. These parameters can be derived directly from detailed physical simulation, or they can be inferred from high-level scene context acting as a prior (\eg classifying a room as a \squote{hallway} provides strong clues about its likely acoustics) \cite{kim_2019}. Relying on priors is computationally efficient but may lack the specificity of direct estimation.

\subsection{\oursystem's Position in the Design Space}
Informed by this analysis, our hybrid approach carves a specific, balanced path through the design space to maximize perceptual realism on resource-constrained hardware. \oursystem's design is a synergistic fusion of pragmatic choices across the three dimensions:
\begin{itemize}
\item For Geometry, \oursystem adopts a hybrid technique: it uses real-time planar surface detection to construct a localized shoebox model around the listener. This approach leverages the efficiency of a shoebox representation while grounding it in the actual dimensions of the user's immediate surroundings, offering a more dynamic and accurate solution than a static, global shoebox.
\item \oursystem employs broad classification, moving beyond geometry-only approaches to provide significant perceptual gains without attempting impractical, fine-grained estimation.
\item \oursystem derives its Acoustic Properties by combining parameters estimated from the specific geometric and material properties with high-level scene context, using the latter as a robust prior to guide the final synthesis.
\end{itemize}
This integrated design enables the on-the-fly, perceptually-driven audio parameterization that is central to our work. The following section details the end-to-end system that implements these choices.
\section{\oursystem}
\label{sec:system}

\begin{figure}
\includegraphics[width=\columnwidth]{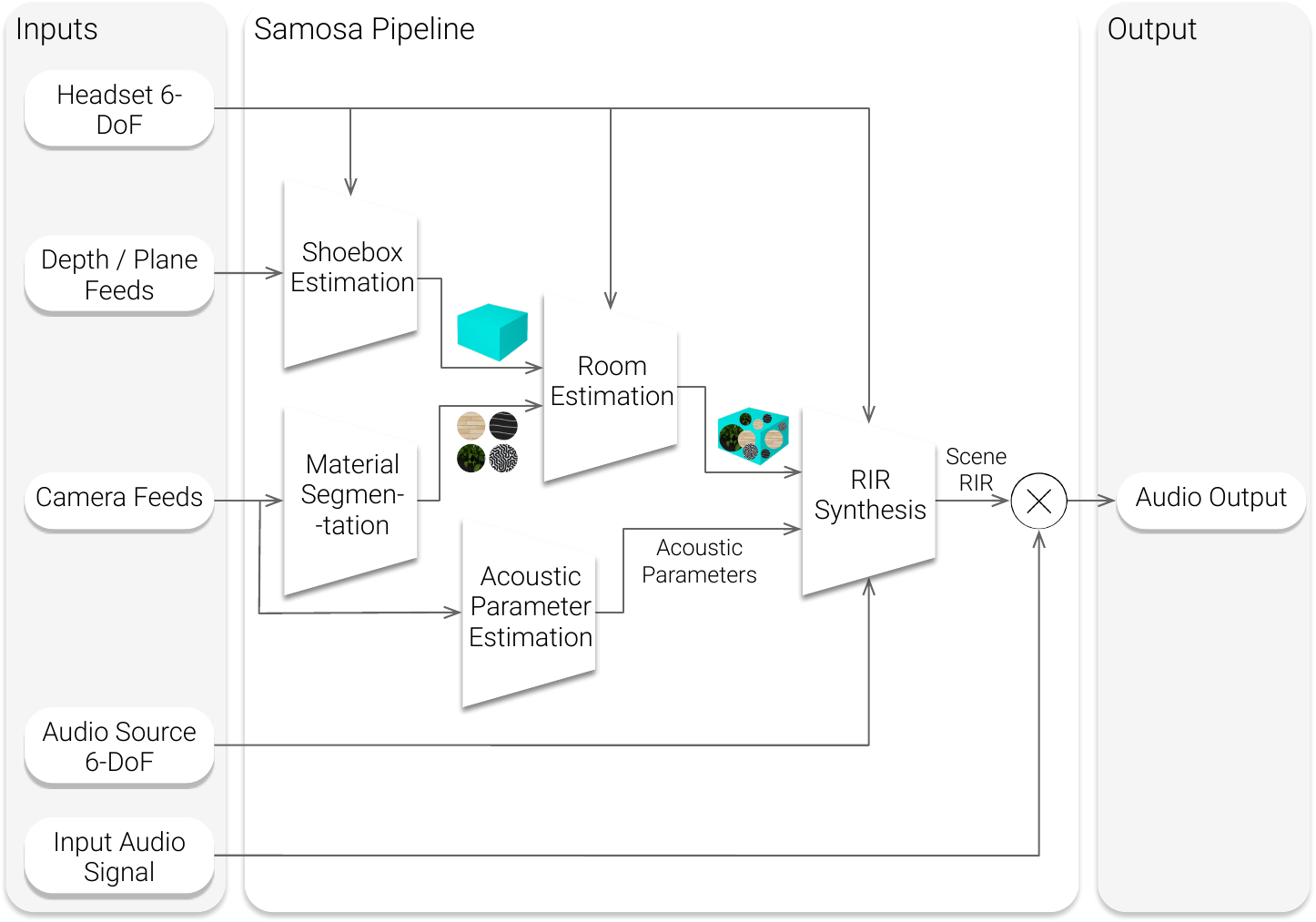}
\vspace{-15pt}
\caption{Overview of the \oursystem pipeline. Inputs such as Depth/Plane Feeds and Camera Feeds are processed by three parallel perception modules: Shoebox Estimation, Material Segmentation, and Acoustic Parameter Estimation. The geometry and material outputs are fused in the Room Estimation step. This combined room model, along with the estimated Acoustic Parameters, Headset 6-DoF, and Audio Source 6-DoF, all inform the RIR Synthesis module. The resulting Scene RIR is then convolved with the Input Audio Signal to produce the final Audio Output.}
\Description{}
\label{fig:pipeline}
\end{figure}

This section details the architecture and implementation of \oursystem, our proposed pipeline for real-time, scene-aware spatial audio rendering. As illustrated in Figure~\ref{fig:pipeline}, the system processes sensor inputs through parallel perception modules for Shoebox Estimation (Sec.~\ref{sec:shoebox_estimation}) and Material Segmentation (Sec.~\ref{sec:material_segmentation}). These streams are fused in the Room Estimation step (Sec.~\ref{sec:room_estimation}) to create a scene model. This model, along with parameters from the Acoustic Parameter Estimation module (Sec.~\ref{sec:acoustic_parameter_estimation}), configures the final RIR Synthesis stage (Sec.~\ref{sec:rir_synthesis}). The subsequent subsections describe each component in detail.

\subsection{Shoebox Estimation}
\label{sec:shoebox_estimation}
\begin{figure}
\includegraphics[width=\columnwidth]{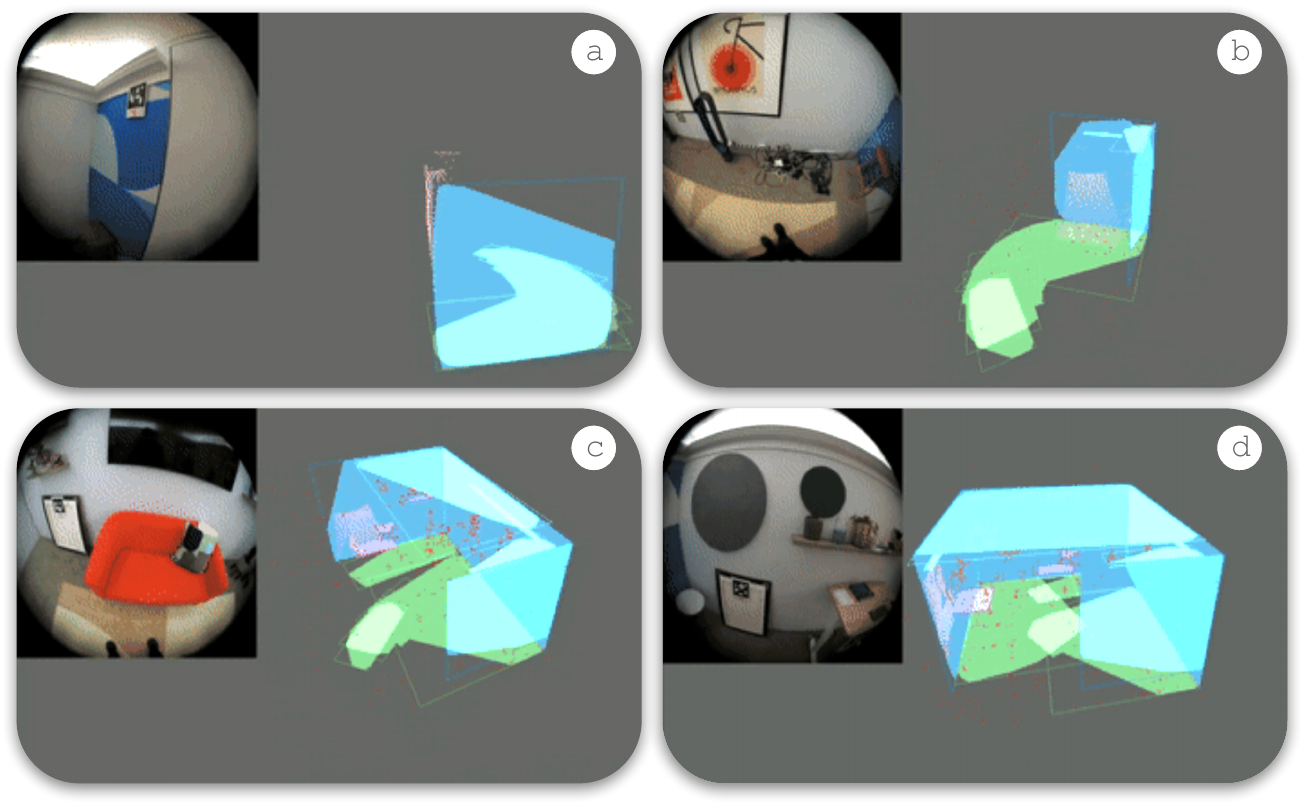}
\caption{Visualization of \oursystem's dynamic Shoebox Estimation. Figures (a) through (d) illustrate how the estimated shoebox room geometry is updated over time as a user explores a new scene.}
\label{fig:geometry}
\end{figure}

Real-time audio rendering requires efficiently processing room geometry, captured via plane estimation from depth feeds. While plane estimation yields detailed surfaces, the resulting irregular polygonal meshes are computationally prohibitive for real-time geometric acoustics. To address this, we employ a shoebox approximation to generate a simplified, axis-aligned cuboid representation of the room from the primary structural planes.

This simplification necessitates mapping dynamic entities—audio sources and the listener—into the shoebox. Our mapping strategy prioritizes preserving key perceptual cues: surface proximity and relative listener-to-source orientation. We calculate new coordinates $(x', y', z')$ within the shoebox such that distances to the three nearest orthogonal walls match the original distances to the corresponding real-world surfaces. To preserve relative orientation, if the mapping changes the source-listener vector from $\vec{v}_{orig}$ to $\vec{v}_{new}$, we apply the minimal rotation matrix $\mathbf{R}$ ($\mathbf{R}\vec{v}_{orig} = \vec{v}_{new}$) to the listener's original orientation. This approach preserves geometric cues influencing primary reflections and perceived source direction.

\subsection{Material Segmentation}
\label{sec:material_segmentation}

\begin{figure}
\includegraphics[width=\columnwidth]{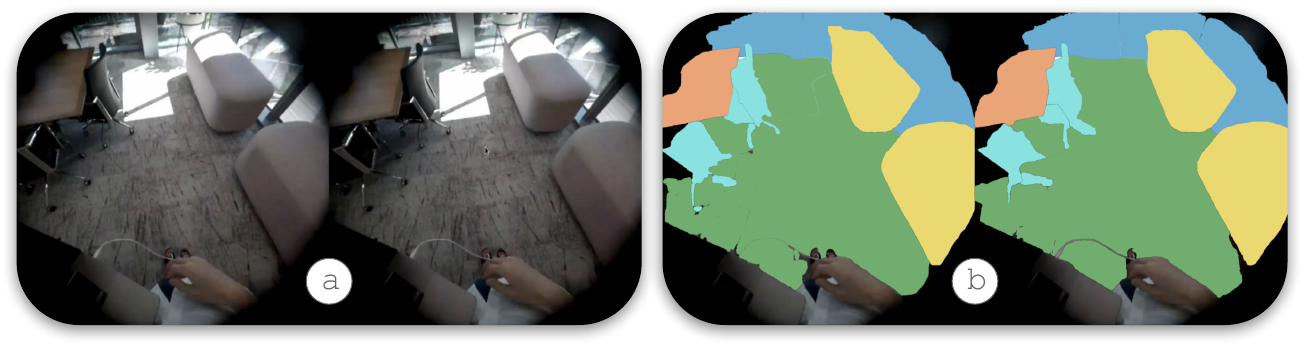}
\caption{Visualization of \oursystem's Material Segmentation. (a) Input binocular egocentric view. (b) Corresponding output segmentation map where pixels are classified into material categories.}
\label{fig:material}
\end{figure}
The material segmentation module processes camera images to predict pixel-wise material labels. We use a transfer learning approach, adapting a pre-trained semantic segmentation model by adding a dedicated prediction head for material classification.

\subsubsection{Model Architecture}
We use a multi-head DeepLabv3+ architecture \cite{chen_2018} with a MobileNetV2 backbone \cite{MobileNetV2}, ensuring efficient feature reuse with other on-device semantic tasks. A dedicated head is added and trained specifically for predicting 9 relevant material classes.


The core architecture consists of:
\begin{itemize}
    \item \textbf{Backbone:} A MobileNetV2 feature extractor.
    \item \textbf{Decoder:} An Atrous Spatial Pyramid Pooling (ASPP) module \cite{chen_2018} processes the backbone features at multiple scales to capture contextual information, configured with depthwise separable convolutions for efficiency.
    \item \textbf{Material Segmentation Head:} \change{Following the DeepLabv3+ design, low-level features (from MobileNetV2) are fused with the ASPP output before passing through convolutional layers (2 layers, 256 filters each, depthwise separable) and a final 1x1 convolution to predict segmentation logits. In our multi-head setup, a dedicated head is added specifically for predicting material classes, parallel to any existing semantic segmentation head.}
\end{itemize}

\subsubsection{Training}
We trained the material segmentation head using a transfer learning strategy, freezing the pre-trained MobileNetV2 backbone and updating only the parameters of the new head. The key training details and hyperparameters are as follows:

\begin{itemize}
\item \textbf{Dataset and Annotation:} \change{We started with the COCO-Stuff dataset \cite{cocostuff}. To generate training data for our 9 target material classes, we used an automated labeling pipeline. First, the Segment Anything Model (SAM) \cite{SAM} generated high-quality instance masks. Then, following the method of Peng et al. \cite{peng_2023}, a Vision-Language Model (VLM) automatically assigned a material label to each mask by comparing image and text embeddings.}
\item \textbf{Input Processing:} Images were resized to 320x320 pixels. We applied standard data augmentation (random horizontal flipping and scaling from 0.5x to 2.0x) and normalized pixel values using ImageNet statistics \cite{imagenet}.

\item \textbf{Loss Function:} A standard cross-entropy loss was used, with top-k pixel selection (k=0.25) to focus on harder examples. L2 weight decay (4.0e-5) was applied for regularization.

\item \textbf{Optimizer:} We used Stochastic Gradient Descent (SGD) with a momentum of 0.9.

\item \textbf{Learning Rate:} A cosine decay schedule was employed with an initial learning rate of 0.08, following a 770-step linear warmup phase.
\end{itemize}

This training regime adapts the shared features to identify common indoor materials relevant for acoustic modeling.

\subsection{Room Estimation}
\label{sec:room_estimation}

\change{The Room Estimation module, shown in Figure~\ref{fig:pipeline}, fuses the outputs of the geometry and material perception streams to create a unified, room model. This crucial intermediate step associates the 2D material classifications (Sec.\ref{sec:material_segmentation}) with the 3D surfaces of the estimated shoebox model (Sec.\ref{sec:shoebox_estimation}).
The association process projects the 3D shoebox faces into the 2D camera view and aggregates the material segmentation results for each face over time as the user looks around. To create a detailed acoustic profile for each surface, our system moves beyond a single-label assignment. For each of the six shoebox surfaces, it maintains a distribution of up to a maximum of 10 of the most prominent materials detected. For each material in this distribution, it continuously calculates and updates two key metrics:
\begin{itemize}
\item \textbf{Area Ratio:} The proportion of the surface area covered by that material, based on the pixel count in the segmentation map.
\item \textbf{Confidence Level:} An aggregated confidence score derived from the underlying segmentation model's output for the corresponding pixels.
\end{itemize}
The final output for each surface is therefore not a single label, but a rich material distribution (\eg a wall might be represented as 70\% heavy curtain and 30\% glass, each with its own confidence). This detailed profile allows the RIR Synthesis module to compute more accurate, blended acoustic properties for each surface, leading to a more realistic sound environment.
}
\subsection{Acoustic Parameter Estimation}
\label{sec:acoustic_parameter_estimation}

While geometry and materials define a room's structure, they are often insufficient for capturing its global acoustic character. Directly regressing detailed acoustic values like RT60 from visual data is also a complex, unresolved problem. Therefore, \oursystem employs a pragmatic and efficient two-stage approach: first, it performs a high-level visual classification of the environment (\eg \squote{living room}). Second, it maps this classified scene type to a compact vector of pre-optimized acoustic parameters that drives the final audio synthesis.

\subsubsection{Stage 1: Scene Type Classification}
First, a MobileNetV2 classifier categorizes the environment into one of five types relevant to distinct acoustic behaviors: conference room, living room, bedroom, outdoor, or other. This classification operates on RGB frames and was chosen for its on-device efficiency. The model was trained from scratch on the COCO-Stuff dataset \cite{cocostuff}, for which we generated targeted room-type labels using a Gemini-based large language model \cite{gemini} (see Appendix~\ref{appendix:scene_type_classification_prompt} for the prompt).

\change{To ensure robustness against real-world visual variations, the training incorporated extensive data augmentation (\eg blur, crop/resize, color adjustments). The training used the RMSprop optimizer \cite{HintonRMSprop} with an exponential learning rate decay schedule (initial LR 0.0002, decay rate 0.75 per 152 steps, staircase=true), cross-entropy loss with label smoothing (0.1), and L2 weight decay (1.0e-7) for regularization.}

Notably, the training strategy for this classifier was deliberately different from the material head's. Unlike the material head, which used transfer learning with a frozen backbone, the scene classifier was trained from scratch with the backbone unfrozen (\texttt{freeze\_backbone=false}). This approach allows the entire network to learn higher-level feature representations better suited for holistic scene classification, rather than only reusing low-level texture features.

\subsubsection{Stage 2: Mapping Scene Type to Acoustic Parameters}
\change{Each of the five classified scene types is mapped to a corresponding embedding vector of acoustic parameters (\eg values controlling reverberation time, gain, and brightness). These parameter vectors are determined empirically through a one-time, offline optimization process designed to best match real-world acoustics.

We performed a grid search to find the optimal parameter vector for each scene type. The optimization objective was to minimize the Mean Absolute Error (MAE) between the RT60 generated from a candidate parameter vector and ground truth RT60 measurements from a dedicated validation dataset. The resulting embedding vector provides a concise yet perceptually relevant parametric representation of the room's reverberant characteristics. This stage ensures that each classified scene type is associated with a set of parameters that reliably reproduces the acoustic signature of representative real-world spaces, which are then fed to the RIR Synthesis module.}

\subsection{RIR Synthesis}
\label{sec:rir_synthesis}

The RIR Synthesis module is the final stage of the perception pipeline, responsible for generating a complete Room Impulse Response (RIR) that represents the acoustic characteristics of the user's environment. It uses the estimated room properties—the fused room model from the Room Estimation step, the Acoustic Parameters, and the 6-DoF poses of the listener and source—to dynamically configure our audio rendering engine, which leverages the Resonance Audio SDK \cite{resonance}. This module synthesizes the two perceptually crucial components of the RIR: early reflections (ER) and late reverberation (LR).

\subsubsection{Early Reflection Synthesis}
\paragraph{Parameter Calculation}
This stage calculates a blended, frequency-dependent reflection coefficient for each of the six shoebox faces. The process starts with a library of predefined absorption coefficients for our material classes. A weighted average of these coefficients is then computed for each surface, using the area ratios and confidence levels of all detected materials (from Sec.~\ref{sec:room_estimation}) as weights. This yields coefficients that model the composite nature of surfaces and shape the timbre of reflections. A global reflection gain scalar is also derived from the Acoustic Parameters to control the overall reflection level.
\paragraph{Rendering}
The rendering engine uses these properties and the listener's pose to simulate reflection paths via an image source method (ISM) \cite{allen_1979}. This determines the delay, direction, and attenuation for each reflection, forming the early reflection part of the RIR. Reflections are only rendered when the listener is inside the estimated shoebox.

\subsubsection{Late Reverberation Synthesis}
\paragraph{Parameter Calculation}
The calculation of late reverberation parameters is a hybrid process, combining physically-based estimation with learned, top-down refinement.
First, we compute a baseline frequency-dependent RT60 using Eyring's equation \cite{eyring}, which uses the shoebox volume and the room's total average surface absorption calculated from the weighted average of all detected material properties.
Second, this physically-based RT60 is refined using the output from the Acoustic Parameter Estimation module. This module provides a set of learned parameters based on the classified scene type—including reverb gain, a reverb time modulator, and reverb brightness—that adjust the baseline estimate. This hybrid approach grounds the reverberation in the scene's physical properties while allowing learned parameters to tune the result for a more perceptually accurate character.
\paragraph{Rendering}
The rendering engine uses these final, refined properties to generate the late reverberation tail via a spectral reverberation technique \cite{jot_1992}. This process filters an impulse into frequency bands and applies a unique energy decay envelope to each, based on the refined RT60 values, before decorrelating and summing them to produce the late reverberation component of the RIR.

\change{\subsubsection{RIR Composition and Final Audio Rendering}
The synthesized early reflections and late reverberation are combined with a direct path component to form a complete Scene RIR. As shown in Figure~\ref{fig:pipeline}, the final scene-aware spatial audio is then rendered in real-time by convolving this dynamically generated RIR with the input audio signal.
While this description focuses on a single sound source, the RIR synthesis process scales efficiently to multi-source scenarios. This is achieved by computing unique early reflections for each source's RIR based on its position, while reusing a single, shared late reverberation model for the entire scene. This approach maintains plausible spatial cues for all sources without a prohibitive increase in computational cost.}

\subsection{Implementation}
The proposed system was implemented and optimized for deployment on resource-constrained XR hardware. We developed two primary versions: an audio processing pipeline integrated within \ourheadset designed for AR and MR use cases, and a plugin for the Unity engine to facilitate VR development. In this section, we elaborate on the implementation details of the key components of \oursystem.

\subsubsection{Shoebox Estimation} This module achieves position errors below 0.1 meters and orientation errors less than 0.9 radians during tests. For real-time operation on the target XR headset, this module consumes less than 1\% of the CPU resources and approximately 0.48W of power.
\subsubsection{Scene Understanding} Semantic analysis of the environment relies on two core machine learning models: one for material segmentation and another for scene classification. Both models were trained using Quantization-Aware Training (QAT) to optimize their size and inference speed for on-device execution: 
\begin{itemize}
    \item The material segmentation model required 20 hours and 22 minutes to train, utilizing approximately 1.17 CPU cores (GCU), 11.5 GiB of memory, and 8 TPU accelerators ('JellyDonut' TPUs). The resulting optimized model has a size of 2.5MB and contains 33,058,383 parameters.
    \item The scene classification model was trained in 4 hours and 51 minutes using about 0.59 CPU cores (GCU), 11.6 GiB of memory, and 8 JellyDonut TPU accelerators. Its final optimized size is 956KB, encompassing 8,975,715 parameters.
\end{itemize}

\subsubsection{Audio Rendering} The audio rendering pipeline, built upon the Resonance Audio SDK \cite{resonance}, has an end-to-end latency of 58ms and utilizes less than 2\% of the CPU.

\change{In summary, the constituent models of the \oursystem pipeline have a combined memory footprint of approximately 3.5 MB. The full end-to-end system, from sensor input to audio output, delivers real-time, scene-aware acoustic rendering with a low latency of 58 ms while consuming under 3\% of the total on-device computational resources, demonstrating its strong performance and suitability for always-on deployment on XR hardware.}

\section{Application Scenarios}
\label{sec:applications}

\begin{figure*}
  \includegraphics[width=\textwidth]{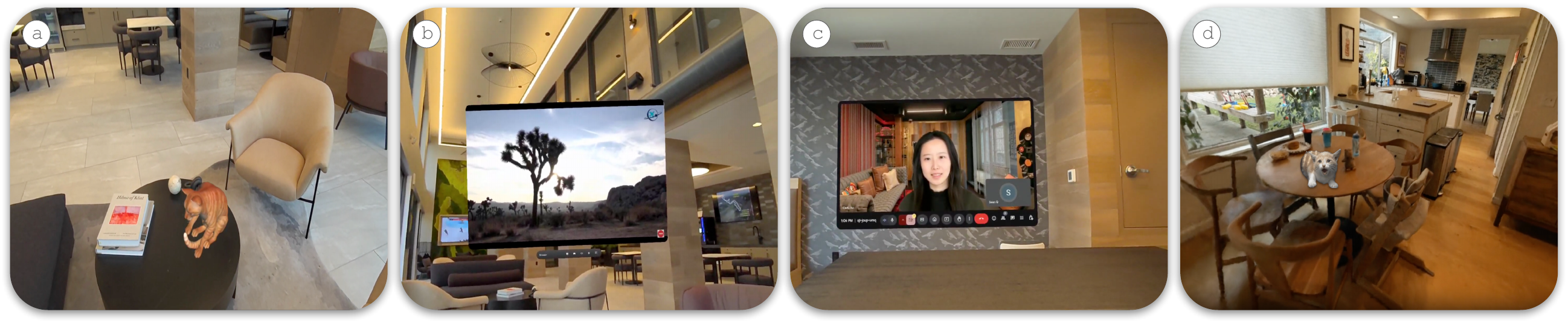}
  \vspace{-15pt}
  \caption{Example application scenarios enabled by \oursystem. Applications illustrated are: (a) Rendering a virtual cat in Video-See-Through (VST) mode with plausible environmental acoustics. (b) Enhancing auditory co-presence in XR video conferencing by matching remote user audio to the local space. (c) Creating an immersive video watching experience where ambient sounds blend with the real room. (d) Realistic rendering of audio sources within a detailed VR scene reconstructed via Zip-NeRF \cite{zipnerf_2023}.}

  \label{fig:applications}
\end{figure*}

As an enabling technology, we envision \oursystem inspiring future XR applications by realistically rendering virtual audio into the user's physical environment. Key scenario include:

\paragraph{Spatially Anchored Audio in AR/VST.}
\oursystem allows virtual sound sources placed in AR or Video-See-Through (VST) views to interact convincingly with the physical surroundings. For instance, users viewing XR assets with VST enabled could hear a virtual music box placed on a real table as if the music originates from that surface, accurately reflecting off nearby walls (Figure \ref{fig:teaser}). Similarly, the sounds from virtual objects like a toy plane flying around the room or a virtual cat (Figure~\ref{fig:applications}a) would be realistically processed based on the room's geometry and materials. This capability enhances AR/VST games and experiences with virtual pets by rendering dynamic audio effects (\eg character sounds, object interactions) that are plausibly occluded or reflected by real-world obstacles \cite{vrpet2021}, leading to more intuitive spatial awareness and believable interactions. 

\textit{Co-presence and Teleportation.}
\oursystem can improve video conferencing and remote collaboration by fostering a greater sense of shared space (Figure~\ref{fig:applications}b). Instead of remote participants' voices sounding disconnected or generic, our scene-aware rendering can make them sound as if they are physically present in the listener's local environment, with their voices reflecting off the actual walls and surfaces of the room. This heightened realism significantly improves the feeling of co-presence and makes interactions feel more natural and engaging, moving towards true virtual teleportation experiences \cite{holoportation_2016}.

\textit{XR Entertainment.}
\oursystem enables deeply immersive entertainment experiences (Figure~\ref{fig:applications}c) by tailoring audio rendering to the user's specific physical space. For example, users could listen to music (\eg from YouTube Music in a web browser) rendered through simulated high-fidelity speakers positioned virtually within their room, with acoustics accurately matching the room's characteristics. Similarly, virtual concerts could acoustically adapt to the listener's living room \cite{metaverse}, enhancing the sense of presence. Furthermore, the dynamic rendering of sound effects (\eg explosions, dialogue) in AR games to plausibly interact with the player's physical surroundings can significantly increase immersion and realism \cite{ar_survey}.

\textit{Immersive VR Scenes.}
\change{Beyond AR/VST, \oursystem also enhances auditory realism in fully virtual environments (Figure~\ref{fig:applications}d). Instead of relying on sensor-based scene perception, it operates on reconstructed 3D scenes in VR. Paired with high-fidelity visual rendering—such as NeRFs \cite{zipnerf_2023} or real-time Gaussian Splatting methods like RadSplat \cite{radsplat_2025}—our system ensures that spatial audio coherently aligns with the detailed virtual world for a more convincing and immersive experience.}


\section{Technical Evaluation}
\label{sec:technical_eval} 

In this section, we evaluate \oursystem's multimodal RIR synthesis capabilities by comparing its output against ground truth acoustic measurements and established baselines using standard objective metrics. We first introduce the datasets, then describe the baseline methods and evaluation metrics, and finally present the quantitative results and our findings.

\subsection{Methods}
\subsubsection{Datasets}
\label{datasets}

 We collected data from 25 diverse scenes using a set of identical, custom-built XR headset prototypes equipped with the Snapdragon XR2+ Gen 2 chipset \cite{qualcomm:xr2plusgen2_brief}. For each scene, we recorded time series data including headset 6DoF poses, corresponding pass-through images, and depth information. The collection covers various environments, including conference rooms with different geometries and diverse home settings. For quantitative evaluation, we captured high-fidelity ground truth RIRs and associated RT60 measurements in a subset of these scenes under controlled conditions, as detailed below. To assess both single-scene RIR estimation accuracy and multi-scene dynamic adaptation, we collected two datasets:

\paragraph{High-Fidelity Single-Scene (HFSS) Dataset}
This dataset comprises detailed scans and high-quality acoustic measurements from 5 distinct rooms. It serves as the primary benchmark for evaluating the fundamental accuracy of \oursystem's simulated RIR against precise, real-world acoustic ground truth.
We first captured the ground truth scene geometries and a 360 image of the room as semantic ground truth. For acoustic ground truth, a high-quality coaxial speaker (Genelec 8341) was placed one-third of the room's floor diagonal away from a corner, facing inwards. Two reference microphones (B\&K 4942) were positioned within the room (one near the center, one near an edge or corner). We measured the position of the speaker and microphones within the room. An exponential sine sweep (equalized from 50Hz to 9kHz) was played through the speaker and recorded via the microphones connected to a B\&K 3670 USB audio interface. RIRs were generated using the Farina method \cite{farina_2000}. Additionally, RT60 measurements were taken at both microphone positions using a B\&K 2250 SPL meter for verification.
Within these same 5 rooms, we also characterized the approximate RIRs generated by the Apple Vision Pro (AVP)'s spatial audio effect during FaceTime calls \cite{avp}. These measurements involved placing an iPhone 10 inches from a B\&K 4227A artificial mouth speaker within an acoustically treated enclosure in one room. A FaceTime call connected this iPhone to an AVP in the target room. The AVP's audio/video output was mirrored and recorded using OBS Studio on a laptop while the exponential sine sweep was played through the artificial mouth. The recorded audio, processed by the AVP's FaceTime reverberation effect, was then used to generate impulse responses following the same procedure.

\textit{Dynamic Multi-Scene (DMS) Dataset.}
This dataset focuses on evaluating \oursystem's dynamic adaptation capabilities. It features scans and audio captures from 20 scenes, primarily consisting of pairs of neighboring spaces. Data was captured under varying scene transitioning conditions, including different types of connecting doors (\eg wooden swing doors, glass doors, sliding doors) and user head movements. We captured similar headset scans and ground truth geometries as described in the \textit{HFSS Dataset}. However, the acoustic ground truth here used exponential sine sweeps played from various commercially available speakers and were recorded using the built-in microphones of the headset. This dataset is crucial for assessing \oursystem's core capability: dynamically adapting its RIR synthesis based on real-time, on-device scene understanding in response to changes in the physical environment, such as opening a door or moving between connected spaces. The diversity in room pairs and configurations tests the robustness of this adaptation mechanism.

\subsubsection{Baselines}
To contextualize \oursystem's performance, we compared its RIR estimation against a spectrum of baselines, from scene-agnostic methods to state-of-the-art offline models.

\begin{enumerate}
\item \textbf{\NonAdaptive Baseline:} Utilized a fixed, generic reverb preset, representing systems with no scene awareness.
\item \textbf{Alternative RIR Estimation Baselines:} We compared against recent learning-based methods:
    \begin{description}
        \item [\textbf{Image2Reverb (Visual-Only):}] This model~\cite{image2reverb_2021} predicts acoustic parameters from a single RGB image.
        \item [\textbf{AV-RIR (Audio-Visual):}] This multimodal approach~\cite{avrir_2024} estimates RIRs from reverberant audio and visual information. 
    \end{description}
    
\item \textbf{\oursystem Ablation Studies:} To quantify the contribution of each component, we evaluated three ablated versions: \textbf{\oursystem-GeoOnly} (using only geometry), \textbf{\oursystem-MatOnly} (using only materials with a canonical room), and \textbf{\oursystem-AEOnly} (using only the acoustic embedding with a default room).

\item \textbf{Commercial System Baseline:} We included the reverberation effect measured from the Apple Vision Pro's FaceTime feature \cite{avp}, as captured and detailed in Section~\ref{datasets}.
\end{enumerate}
\subsubsection{Metrics}
To objectively evaluate the generated RIRs, we computed Reverberation Time (RT60) and Early Decay Time (EDT) based on ISO 3382-1 \cite{ISO3382-1-2009}, averaged across 8 standard octave bands (62.5 Hz to 8 kHz). These metrics are standard for assessing perceived room acoustics \cite{kuttruff_2016}.
\begin{itemize}[leftmargin=*]
\item \textbf{Reverberation Time (RT60)} characterizes the overall perceived duration of reverberance.
\item \textbf{Early Decay Time (EDT)} correlates more closely with the subjective perception of reverberance, as it is sensitive to early reflections influenced by local geometry \cite{kuttruff_2016}.
\end{itemize}
For both metrics, we report Mean Absolute Error (MAE) and Root Mean Squared Error (RMSE) against the ground truth RIRs.

\subsection{Results and Discussion}
\label{subsec:obj_results_discussion}
\subsubsection{Result}
\label{subsubsec:objective_results}
\begin{table*}[htbp]
    \begin{threeparttable}
    \centering
    \begin{tabular}{lcccc}
        \toprule
        Method    & RT60-MAE(s) $\downarrow$ & RT60-RMSE(s) $\downarrow$ & EDT-MAE(s) $\downarrow$ &EDT-RMSE(s) $\downarrow$ \\ 
        \midrule
        Non-Adaptive  & 0.4271 & 0.4645 & 0.2905 & 0.3201 \\ 
        \midrule 
        \oursystem-GeoOnly & 0.3011 & 0.3588 & 0.1711 & 0.2205 \\ 
        \oursystem-MatOnly & 0.3237 & 0.3877 & 0.2896 & 0.3273 \\ 
        \oursystem-AEOnly  & 0.2451 &0.3297 & 0.2254 & 0.2821 \\ 
        \midrule 
        Image2Reverb\tnote{**} & 0.2231 &0.3028 & 0.1942 &0.2234 \\
        AV-RIR\tnote{**}  & 0.1024 &0.1305 &0.0971 &0.1169 \\
        \midrule 
        AVP FaceTime\tnote{*} & 0.1958 & 0.2536 & 0.1542 & 0.2095 \\
        \midrule 
        \textbf{\oursystem (Ours)}& \textbf{0.1761}& \textbf{0.2318}& \textbf{0.1579} & \textbf{0.2054} \\
        \bottomrule
    \end{tabular}
    \caption{Objective evaluation of RIR estimation accuracy.}
    \label{tab:results_objective}
    \vspace{-10pt}
    \begin{tablenotes}[para,flushleft] 
        \item[*] AVP results were obtained only on the HFSS dataset. All other results are averaged over the combined dataset (HFSS + DMS, 25 scenes). 
        \item[**] Offline models like Image2Reverb (687 MB) and AV-RIR (6.1 GB) have large memory footprints. In contrast, our real-time \oursystem model has a \textasciitilde 3.5 MB footprint, an end-to-end latency of 58 ms, and uses <3\% CPU on the target hardware.
    \end{tablenotes}
    \end{threeparttable}
\end{table*}

Table~\ref{tab:results_objective} presents the objective evaluation results. Our proposed method, \oursystem, achieves strong performance across all metrics (\eg 0.1761s RT60-MAE and 0.1579s EDT-MAE).

This performance signifies a substantial improvement over the \NonAdaptive baseline, reducing errors by approximately 36\% to 59\% across the different metrics. Furthermore, comparing the \oursystem against its ablation variants (\SamosaGeoOnly, \SamosaMatOnly, \SamosaAEOnly) confirms that the complete model yields the lowest errors, demonstrating the positive contribution of all integrated components.

When compared to other learned approaches evaluated on the combined dataset, \oursystem outperforms \imagereverb (\eg 0.1761s vs. 0.2231s RT60-MAE). While \avrir achieves the lowest overall error scores (\eg 0.1024s RT60-MAE), it requires significantly higher model complexity: \avrir uses 6.1 GB of memory and \imagereverb uses 687 MB, compared to merely 4 MB for \oursystem, as noted in Table~\ref{tab:results_objective}.

While \textit{AV-RIR} achieves the lowest numerical error, it is a computationally expensive offline model. As detailed in Table~\ref{tab:results_objective}, \textit{AV-RIR} (6.1 GB) and \textit{Image2Reverb} (687 MB) are orders of magnitude larger than \oursystem (\textasciitilde3.5 MB), which operates efficiently in real-time with an end-to-end latency of just 58 ms while using under 3\% of the CPU.

Finally, compared to the commercial \avpfacetime baseline, \oursystem shows roughly 9-10\% lower error for RT60 metrics and performs comparably on EDT metrics (slightly higher MAE, slightly lower RMSE). As indicated in the table notes, this comparison should consider that the \avpfacetime results were obtained only on the HFSS dataset subset, whereas \oursystem's results are averaged over the larger combined HFSS and DMS dataset.

\subsubsection{Findings}
\label{subsec:discussion}
\change{The evaluation demonstrates that our integrated model is effective, substantially outperforming the non-adaptive baseline and its own ablated versions, and achieving accuracy superior to \imagereverb. The performance of the full model compared to its ablated versions highlights the synergistic benefit of integrating geometric, material, and acoustic context for robust RIR synthesis. 

The comparison with \avrir highlights \oursystem's primary advantage: achieving a compelling balance between predictive accuracy and exceptional computational efficiency. Its lightweight nature (\textasciitilde3.5 MB, <3\% CPU) makes it a viable and practical solution for real-time processing on resource-constrained XR devices, where deploying large offline models like \avrir or \imagereverb is infeasible. Furthermore, \oursystem achieves strong objective accuracy even under dynamic, on-the-fly conditions. As our validation started from a fresh system state, these results reflect high performance with only a partial environmental model, confirming the system's efficacy without requiring an exhaustive pre-scan.

In conclusion, the findings position \oursystem as a highly effective approach that balances strong RIR estimation accuracy with the practical demands of on-device XR systems.}

\section{Expert Evaluation}
\label{sec:expert_evaluation}
\begin{figure*}[ht]
  \centering
  \includegraphics[width=\textwidth]{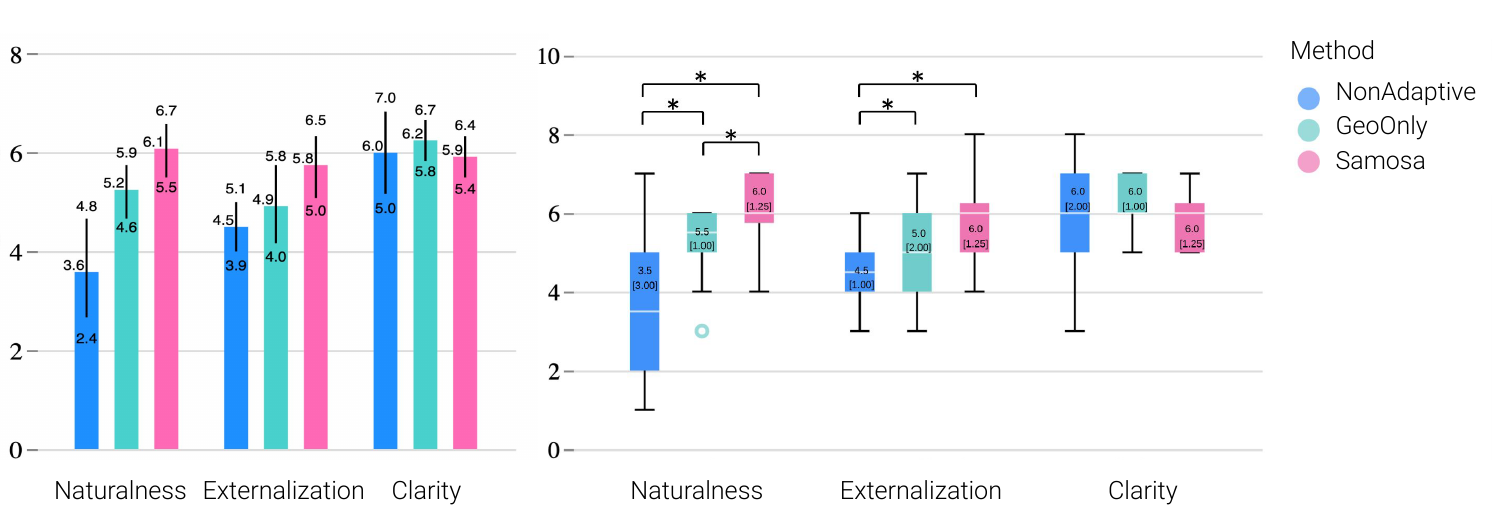}
  \vspace{-25pt}
  \caption{Expert evaluation results comparing \NonAdaptive, \SamosaGeoOnly, and \SamosaOurs audio rendering methods across three perceptual metrics (0-10 scale, higher is better). \textbf{Left:} Mean ratings with 95\% Confidence Interval (CI) error bars. \textbf{Right:} Median ratings with Interquartile Range (IQR = Q3 - Q1) error bars. Statistical significance of pairwise comparisons is detailed in Figure~\ref{fig:expert_eval_stat_sig}.}
  \label{fig:expert_eval_plots}
\end{figure*}

\change{We conducted an expert evaluation to assess how different real-time, on-device rendering configurations affect perceived audio quality in an XR environment.}

\subsection{Participants}

\change{We recruited 12 expert listeners (8 male, 4 female; age range: 29–45, M = 34.2, SD = 4.4) from \ourcompany. The cohort comprised professional musicians, audio engineers, spatial audio researchers, XR sound designers, and specialized audio test engineers, with an average of 11.3 years of professional experience in audio (SD = 4.0). Following best practices for subjective audio assessment (\eg ITU-R BS.1534-3) \cite{itu_2003, itu_2015, aalto}, participants were selected for their heightened sensitivity to subtle acoustic differences and their ability to provide articulate feedback. All participants reported normal hearing.}

\subsection{Apparatus}
\subsubsection{Hardware}
\change{The study utilized four identical custom-built XR headset prototypes equipped with the Snapdragon XR2+ Gen 2 chipset \cite{qualcomm:xr2plusgen2_brief}. Audio stimuli were delivered through the integrated near-ear spatial speakers on the headsets. The evaluation took place in a controlled lab environment composed of three interconnected rooms, each with distinct dimensions and acoustic characteristics: a large open area ($\sim$20m x 15m x 4m), a medium-sized conference room ($\sim$4m x 7m x 3m), and a small meeting room ($\sim$3m x 3m x 3m).}

\subsubsection{Software}
\label{subsubsec:software_apparatus}

The evaluation software was designed to compare different real-time, on-device rendering methods within a hardware-controlled setting to isolate algorithmic performance. Each headset ran one of three candidate builds:

\begin{enumerate}
    \item \textbf{\NonAdaptive Baseline:} A standard rendering configuration without dynamic scene adaptation, serving as a control.
    \item \textbf{\SamosaGeoOnly:} Our system adapted using only estimated scene geometry, isolating the effect of geometric awareness.
    \item \textbf{\SamosaOurs:} The full proposed system incorporating geometry, materials, and acoustic context estimation.
\end{enumerate}

\change{Our baseline selection focused on methods suitable for real-time, interactive XR on standalone headsets. Consequently, we excluded computationally intensive offline models like Image2Reverb \cite{image2reverb_2021} and AV-RIR \cite{avrir_2024}, as their high computational demands preclude real-time execution on our target hardware (see Sec. \ref{subsec:obj_results_discussion}). We also excluded proprietary commercial systems such as the Apple Vision Pro \cite{avp}. This was due to several factors that would confound a controlled comparison: the lack of feature availability for evaluation within a third-party application, the introduction of hardware-confounding variables (e.g., speaker quality), and the inability to replicate its closed-source algorithms.}


\subsection{Procedure}

The study employed a within-subjects, blind design where each participant evaluated all three candidate builds, presented in a randomized order to mitigate order effects. Upon arrival, participants gave informed consent and received standardized instructions.

For each build, participants enabled video pass-through and were instructed to walk naturally through the lab spaces while listening to three types of stimuli:
\begin{enumerate}
    \item \textbf{System sounds:} Representative UI interaction sounds (\eg button clicks, notifications).
    \item \textbf{Human speech:} A 30-second clip from a video containing clear, anechoic speech.
    \item \textbf{Music:} A 30-second diverse, dynamic musical excerpt.
\end{enumerate}
\change{The stimuli were chosen to represent common XR audio types, in line with established evaluation guidelines \cite{itu_2015, aalto}. Each evaluation started from a fresh system state, without requiring an pre-scan. After testing each build, participants rated various perceptual qualities on an evaluation form and could revisit methods before finalizing their scores. We collected both quantitative ratings and qualitative feedback. }

\subsubsection{Quantitative Measures:} We focused on three key attributes of spatial audio realism and quality: \Naturalness, \Externalization, and \Clarity \cite{rumsey_2002, lindau_2012, blauert_1997}. For each build and stimulus type, participants rated these attributes on a 0-10 continuous Likert-style scale anchored with labels (\eg 0 = ``Very Unnatural'' / ``Inside Head'' / ``Very Unclear'', 10 = ``Very Natural'' / ``Outside Head'' / ``Very Clear'') based on the following definitions:

\begin{itemize}
\item \Naturalness: How closely the audio resembled sounds occurring naturally within the physical surroundings, free from artifacts.
\item \Clarity: How easy the audio was to perceive and understand, ensuring distinct elements were clearly discernible.
\item \Externalization: The degree to which sounds seemed to originate from external sources in the environment, rather than inside the listener's head.
\end{itemize}

\subsubsection{Qualitative Measures:} Following the ratings, a semi-structured interview was conducted. Participants elaborated on their ratings, preferences, and any perceived audio artifacts. The entire session took approximately 45 minutes per participant. Interview notes were thematically analyzed \cite{braun_2006}, with support from Gemini 2.5 Pro (Experimental) \cite{gemini-2.5-pro-exp}, to identify recurring themes and gather detailed explanations for the quantitative ratings (see Appendix ~\ref{appendix:expert_interview_prompt}).

\subsection{Results and Discussion}
\subsubsection{Results}
\label{sec:expert_results}
\begin{figure}[t]
  \centering
  \includegraphics[width=\columnwidth]{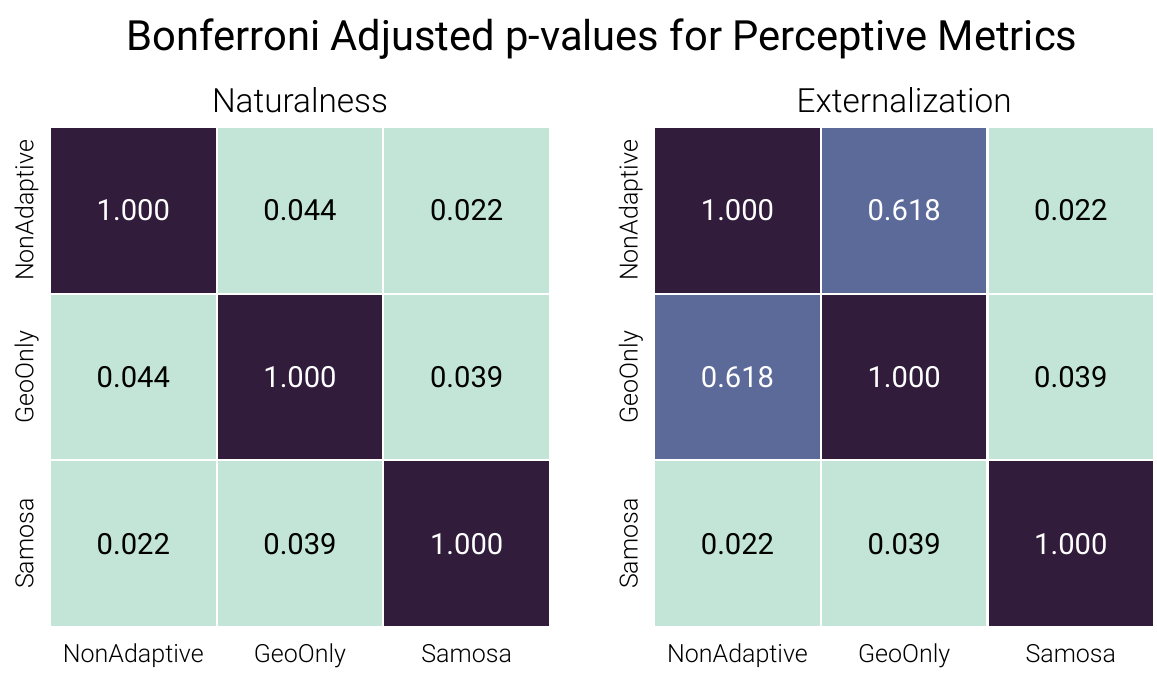}
  \vspace{-15pt}
  \caption{Visualization of Bonferroni-adjusted p-values ($p_{adj}$) from post-hoc pairwise Wilcoxon signed-rank tests comparing the three rendering methods (\NonAdaptive, \SamosaGeoOnly, \SamosaOurs) for the \Naturalness{} and \Externalization{} metrics. Comparisons where $p_{adj} < 0.05$ indicate statistically significant differences.}
  \label{fig:expert_eval_stat_sig}
  \vspace{-10pt}
\end{figure}

We analyzed expert ratings for \Naturalness, \Externalization, and \Clarity using non-parametric tests, appropriate for the ordinal Likert scale data and our sample size. Friedman tests were conducted for each metric to assess overall differences among the three rendering methods (\NonAdaptive, \SamosaGeoOnly, \SamosaOurs). Where significant differences were found, post-hoc pairwise comparisons were performed using Wilcoxon signed-rank tests with Bonferroni correction to identify specific differences between methods. An alpha level of 0.05 was maintained for all statistical tests. Descriptive statistics (mean/95\% CI and median/IQR) are presented in Figure~\ref{fig:expert_eval_plots}, and the statistical significance of pairwise comparisons is visualized in Figure~\ref{fig:expert_eval_stat_sig}.

\paragraph{\Naturalness}
A Friedman test revealed a statistically significant difference in \Naturalness{} ratings across the three methods, $\chi^2(2) = 13.0$, $p = 0.0015$. Post-hoc Wilcoxon signed-rank tests with Bonferroni correction (pairwise significance detailed in Figure~\ref{fig:expert_eval_stat_sig}) indicated that \SamosaOurs{} was rated significantly higher than both \NonAdaptive{} ($p_{adj} = 0.022$) and \SamosaGeoOnly{} ($p_{adj} = 0.039$). Furthermore, \SamosaGeoOnly{} was rated significantly higher than the \NonAdaptive{} baseline ($p_{adj} = 0.044$).

\paragraph{\Externalization}
A Friedman test showed a statistically significant difference in \Externalization ratings across the methods, $\chi^2(2) = 14.6$, $p = 0.0007$. Post-hoc Wilcoxon signed-rank tests with Bonferroni correction (detailed in Figure~\ref{fig:expert_eval_stat_sig}) revealed that both \SamosaOurs{} ($p_{adj} = 0.007$) and \SamosaGeoOnly{} ($p_{adj} = 0.039$) were rated significantly higher than the \NonAdaptive{} baseline. No significant difference was found between \SamosaOurs{} and \SamosaGeoOnly{} ($p_{adj} = 0.618$).

\paragraph{\Clarity}
A Friedman test indicated no statistically significant difference for \Clarity across the three methods, $\chi^2(2) = 1.1$, $p = 0.592$. Consequently, post-hoc pairwise comparisons were not performed for this metric, and it is omitted from the significance visualization in Figure~\ref{fig:expert_eval_stat_sig}.

\subsubsection{Discussion}
\label{sec:expert_discussion}
The expert evaluation demonstrates that \oursystem significantly enhances perceived realism and immersion. Statistical analysis shows \oursystem's improvement in both \Naturalness and \Externalization without compromising audio \Clarity.

The progressive improvement in \Naturalness—from \NonAdaptive to \SamosaGeoOnly to \SamosaOurs—suggests that while geometric adaptation is beneficial, our full model provides a further, significant benefit. This was supported by qualitative feedback, with one expert (P9) attributing this to \oursystem's \dquote{smoother reverberant decay,} indicating a more complete environmental model.

For \Externalization, both \oursystem and \SamosaGeoOnly significantly outperformed the \NonAdaptive baseline, confirming their effectiveness at mitigating the \dquote{in-head localization} issue. The lack of a significant difference between the two adaptive methods suggests that scene geometry is the primary driver for externalization in the tested scenarios.

Crucially, the improvements in realism did not degrade audio \Clarity. We included this metric as a safeguard metric and found no significant difference between conditions ($p \approx 0.59$). This confirms that \oursystem's perceptual gains were not achieved at the cost of intelligibility. Participant comments supported this, with one expert stating the \dquote{differences are small} (P3), and another found all methods \dquote{reasonably clear} (P8).

In summary, the evaluation confirms that \oursystem significantly improves \Naturalness and \Externalization over baseline methods. It was rated as more natural than a geometry-only approach, all while preserving audio \Clarity. These results highlight \oursystem's effectiveness in balancing enhanced realism with core usability for on-device spatial audio rendering.

\subsection{Findings}
\label{subsec:insights}

\change{Based on the promising results of our technical and perceptual evaluations, we derive three key insights for the field of real-time audio rendering for XR, opening up potential new directions for future research.

\subsubsection{Synergistic Multimodal Scene Representation} We demonstrate the efficacy of a synergistic representation, which inspires future work to not only advance individual components (e.g., incorporating event-based geometry, learning continuous material properties) but also develop advanced multimodal pipelines that address research questions beyond audio rendering \cite{du_2020}.

\subsubsection{Efficient Acoustic Calibration via Scene Priors.} A crucial insight is the efficiency of using multimodal scene understanding as a strong prior for acoustic calibration. Our modular approach achieves a perceptual quality comparable or superior to monolithic, end-to-end models but at a fraction of the computational cost. This finding illuminates a research path toward democratizing high-fidelity audio on resource-constrained devices by shifting focus from costly end-to-end inference to efficient, prior-driven adaptation.

\subsubsection{Human-Centered Evaluation for Dynamic Systems.} We propose and validate an evaluation framework tailored for adaptive audio systems. By assessing performance dynamically from a \dquote{fresh state}—without requiring a full environmental pre-scan—our methodology more accurately reflects real-world usage. This approach guides future research in quantifying the real-time perceptual thresholds of  acoustic realism and measuring how audio realism influences user behavior in interactive XR tasks. }
\section{Limitations and Future Work}

Our work presents \oursystem, a novel approach to real-time, scene-aware spatial audio. In this section, we acknowledge current limitations, and identify promising directions for future research.

\subsection{Limitations}
\label{subsec:limitations}
Despite its promising results, our work has several limitations:

\subsubsection{Environmental and Acoustic Generalizability.} Our system's training and design are focused on common indoor environments. Consequently, its performance may degrade in acoustically complex outdoor or industrial spaces not well-represented in our datasets, as both the material segmentation and acoustic classification rely on a predefined set of classes.

\subsubsection{Parametric Audio Representation.} While computationally efficient, the parametric RIR representation used (based on the Resonance Audio SDK) may not capture the full acoustic complexity of the real world, particularly for rooms with highly irregular geometries or intricate diffraction patterns.

\subsubsection{Evaluation Scope and Participant Bias.} 
\change{Our perceptual evaluation, while rigorous, has scope limitations. The findings were derived from audio experts within a single organization, using a curated set of stimuli in a controlled lab. This introduces two potential biases: the results may not generalize to non-expert users in diverse real-world scenarios, and the shared institutional culture of the participants may influence their feedback.}

\subsubsection{Reliance on Non-Acoustic Sensing}
\change{Our system infers acoustics purely from non-acoustic sensors (visual, depth), a deliberate design choice that avoids the significant drawbacks of direct acoustic sensing paradigms in consumer XR:
\begin{itemize}
\item \textbf{Active Sensing:} Requiring the system to emit sounds (\eg audible sweeps or clicks) to measure a response can be disruptive and break user immersion.
\item \textbf{Passive Sensing:} Continuously analyzing ambient environmental sound raises substantial privacy concerns, creates a persistent power drain on mobile hardware, and struggles to generalize across the diverse, uncalibrated microphone arrays on consumer devices.
\end{itemize}
While our non-acoustic approach circumvents these issues, it forgoes the potential benefits of ground-truth data from direct acoustic measurements.}

\subsection{Future Work}
Future research can directly address these limitations and extend the system's capabilities:

\subsubsection{Advanced Acoustic Modeling} To move beyond parametric constraints, future work could explore learning a richer, latent representation of RIRs directly from multimodal inputs. Investigating end-to-end architectures that predict acoustic parameters or even full RIRs from sensor data is another promising, albeit data-intensive, direction. Furthermore, leveraging large multimodal models (LMMs) with novel RIR tokenization strategies could unlock new generative capabilities.

\subsubsection{Broader and \dquote{In-the-Wild} Evaluation} To bolster the generalizability of our findings, future studies should involve non-expert participants in more diverse, \dquote{in-the-wild} settings. Employing a wider variety of dynamic and interactive audio content would also strengthen the validation of the system's real-world effectiveness.

\subsubsection{Dynamic Adaptation and Scene Completeness} \change{A key avenue for future research is the system's dynamic adaptation process. As \oursystem operates \dquote{on-the-fly} without a pre-scan, a focused study could investigate the impact of scene completeness on perceptual quality. This could explore how interactive guidance or different user exploration patterns affect the build-up of the environmental model and the resulting audio experience, providing critical insights into the relationship between partial scene knowledge and acoustic realism.
}

\subsubsection{Finer-Grained Acoustic Interactions} The system could be extended to model more subtle acoustic phenomena. Explicitly modeling diffraction around objects and incorporating semantic-based sound occlusion would further enhance realism for demanding applications in gaming, simulation, and virtual production.

\subsubsection{Hybrid Visual-Acoustic Sensing}
\change{A powerful future direction is to design an intelligent hybrid sensing system that incorporates acoustic data while mitigating the issues we identified. This could manifest in two ways:
\begin{itemize}
\item \textbf{Intelligent Active Probing:} The visually-derived environmental model could act as a strong prior to guide a minimal number of non-disruptive active probes (\eg using soft, in-world sounds that feel diegetic to the experience) only when necessary to resolve acoustic ambiguity.
\item \textbf{Opportunistic Passive Sensing:} Instead of being \dquote{always on}, the system could be triggered to briefly analyze ambient sound only when the visual pipeline detects a significant environmental change (\eg the user entering a new room).
\end{itemize}
Such hybrid approaches could enhance acoustic fidelity by incorporating ground-truth data while carefully managing the real-world trade-offs of disruption, power consumption, and user privacy.}
\section{Conclusion}

\change{In this paper, we presented \oursystem, a novel system for real-time, scene-aware RIR synthesis tailored for immersive XR applications. Inspired by prior work in room acoustics and scene understanding, we analyzed the design space for XR acoustics, identifying geometry, material properties, and acoustic parameters as key representations influencing audio rendering. Based on this analysis, \oursystem integrates modules for real-time geometric approximation, material estimation, and acoustic parameter estimation to drive an efficient, adaptive RIR synthesis process. We evaluated \oursystem through both objective acoustic metrics (\rt, \edt) and perceptual assessments (\Naturalness, \Externalization, \Clarity) conducted with audio experts. Our results demonstrate that our approach can achieve high acoustic realism on resource-constrained XR devices. We believe \oursystem represents a significant step towards more immersive and believable XR experiences and opens up promising avenues for future research in enhancing audio rendering realism within dynamic virtual environments.}

\begin{acks}
We thank Dr. Shichen Liu for assistance with image editing in figure preparation.
\end{acks}

\bibliographystyle{ACM-Reference-Format}
\bibliography{sections/references}

\appendix
\newpage
\onecolumn
\section{LLM Prompts}

\subsection{Expert Interview Comment Analysis Prompt}
\label{appendix:expert_interview_prompt}
\begin{lstlisting}
**Goal:** Analyze the qualitative comments collected from an expert evaluation study, focusing on identifying recurring themes, understanding the reasoning behind quantitative ratings, capturing detailed descriptions of perceived realism, and noting any reported artifacts or usability concerns.

**Context of the Study:**
* **Participants:** 12 experts in spatial audio / XR.
* **Conditions Compared:**
    * Condition 1: Non-Adaptive Baseline ('NonAdaptive' in data)
    * Condition 2: Samosa-GeoOnly ('GeoOnly' in data)
    * Condition 3: Samosa (Ours) ('Samosa' in data)
* **Quantitative Metrics Rated:** Naturalness/Realism, Externalization, Clarity (0-10 scale, higher is generally better).
* **Task:** Participants experienced each condition in an XR headset while walking through different rooms (dynamic evaluation) and rated them, followed by a verbal interview where these comments were collected.
* **(Optional) Key Quantitative Findings Summary:** [Provide a brief 1-2 sentence summary if available."]

**Provided Input (Expert Comments):**
```text
[expert comments input]
```
\end{lstlisting}

\subsection{Scene Type Classification: Baseline Prompt}
\label{appendix:scene_type_classification_prompt}
\begin{lstlisting}
System: You are an AI assistant tasked with predicting the type of room given an image. You can only choose one of the classes in conference room, living room, bedroom, outdoor, other. Must return a single world answer.

User: [Image Input] What room is this?
\end{lstlisting}

\subsection{Scene Type Classification: Class Description Style Prompt}
\begin{lstlisting}
System: You are a specialized Image Classification Agent. Your ONLY function is to analyze an image and determine its room type.
You MUST select exactly ONE category from the following list:
- conference_room
- living_room
- bedroom
- outdoor
- other

Carefully examine the image provided by the user.
- Use 'conference_room' for rooms designed for meetings (large table, multiple chairs, presentation equipment).
- Use 'living_room' for common residential relaxation areas (sofas, coffee tables, TV).
- Use 'bedroom' for rooms primarily for sleeping (bed is prominent).
- Use 'outdoor' for any exterior space.
- Use 'other' for any indoor space that does not clearly fit the first three categories (e.g., kitchen, bathroom, office, hallway, gym).

Your response MUST be a single word chosen precisely from the list above. Do NOT include any other text, explanation, punctuation, or formatting.

User: [Image Input] What room is this?
\end{lstlisting}

\subsection{Scene Type Classification: Chain-of-Thought Style Prompt}
\begin{lstlisting}
System: You are an AI assistant classifying room types from images.
Your task is to analyze the image and output a single-word classification.
Allowed categories: conference_room, living_room, bedroom, outdoor, other.

Follow these steps internally:
1. Identify key objects and the overall environment in the image.
2. Compare these features against the definitions:
    - conference_room: Meeting space, large table, chairs, possibly screens.
    - living_room: Relaxation space, sofas, coffee table, TV typical.
    - bedroom: Sleeping space, bed prominent.
    - outdoor: Exterior space.
    - other: Any other indoor space (kitchen, bath, office, etc.).
3. Select the single best-fitting category from the allowed list.
4. Output ONLY that single word.

Your final response must contain nothing but the chosen category word. No explanations, no extra words.

User: [Image Input] What room is this?
\end{lstlisting}

\end{document}